\documentclass[twocolumn]{cinc}
\usepackage[utf8]{inputenc}
\usepackage{graphicx}
\usepackage{amsmath} 
\usepackage{wrapfig}
\usepackage{multirow}
\usepackage{float}
\usepackage{subfig}
\usepackage{svg}
\usepackage{caption}
\usepackage{xurl}
\usepackage{algorithm}
\usepackage{program}
\usepackage{booktabs}

\usepackage{enumitem}
\setlist[enumerate,1]{label= (\arabic*),leftmargin=*,align=right}

\begin{document}
\title{Robust peak detection for photoplethysmography signal analysis}

\author{Márton Á. Goda$^{1}$, Peter H. Charlton$^{2}$, Joachim A. Behar$^{1}$\\
 \ \\
$^1$Faculty of Biomedical Engineering, Technion, Israel Institute of Technology, Haifa, Israel
 \ \\
$^2$University of Cambridge, Cambridge, United Kingdom}

\maketitle

\begin{abstract}

Efficient and accurate evaluation of long-term photoplethysmography (PPG) recordings is essential for both clinical assessments and consumer products.
In 2021, the top opensource peak detectors were benchmarked on the Multi-Ethnic Study of Atherosclerosis (MESA) database consisting of polysomnography (PSG) recordings and continuous sleep PPG data, where the Automatic Beat Detector (Aboy) had the best accuracy. This work presents Aboy++, an improved version of the original Aboy beat detector. The algorithm was evaluated on 100 adult PPG recordings from the MESA database, which contains more than 4.25 million reference beats. Aboy++ achieved an F1-score of 85.5\%, compared to 80.99\% for the original Aboy peak detector. On average, Aboy++ processed a 1 hour-long recording in less than 2 seconds. This is compared to 115 seconds (i.e., over 57-times longer) for the open-source implementation of the original Aboy peak detector. This study demonstrated the importance of developing robust algorithms like Aboy++ to improve PPG data analysis and clinical outcomes. Overall, Aboy++ is a reliable tool for evaluating long-term wearable PPG measurements in clinical and consumer contexts. The open-source algorithm is available on the physiozoo.com website (on publication of this proceeding).
\end{abstract}

\section{Introduction}

Photoplethysmography (PPG) is an optical technic widely used in clinical and commercial settings to detect volumetric variations in blood circulation in the tissue microvascular bed. PPG is typically performed in the fingertip and offers a low-cost and convenient assessment of various physiological systems, including the cardiovascular, respiratory and autonomic nervous systems \cite{Allen_2007}. Its main application fields include heart rate (HR) measurement \cite{Temko_2017} and atrial fibrillation detection \cite{Mannhart_2023}. Potential applications of PPG include sleep staging \cite{Kotzen_2022}, obstructive sleep apnea screening \cite{Behar_2014_Sleepap},  and blood pressure estimation \cite{Mukkamala_2022}. Real-time PPG assessment is particularly important in clinical settings, where it is extensively used for oxygen saturation and heart rate monitoring. In recent years, the popularity of PPG has grown significantly due to the development of consumer devices such as smartwatches \cite{Charlton_2022wearable}, PPG rings \cite{Sel_2023}, sports belts \cite{Kuwahara_2019} and in-ear PPG devices \cite{Davies_2022}.

In 2022, Kotzen \cite{Kotzen_2022_MSc} reported on a benchmarking study of several top open-source PPG peak detectors, including Aboy \cite{Aboy2005}, AMPD \cite{Scholkmann_2012}, Pulses \cite{Lazaro_2014}, Bishop \cite{Bishop_2018}, COPPG \cite{Orphanidou_2015} and HeartPy \cite{Gent_2019}. The benchmarking was performed on the MESA database containing 19,998 hours of continuous PPG data. Aboy \cite{Aboy2005} demonstrated the highest performance. In 2022, Charlton \textit{et al.} benchmarked 16 peak detectors on 8 datasets \cite{charlton_2022detecting}. The \textit{Aboy} peak detector was amongst the best-performing devices. The present study focused on improving the performance and reducing computational complexity of the Aboy peak detector.

%%%%%%%%%%

\section{Peak Detection}

\subsection{\textit{Aboy} peak detector}

In the case of the original \textit{Aboy} algorithm, the PPG signal is heavily filtered to preserve frequencies near an initial heart rate estimate \cite{Aboy2005}. The PPG signal is segmented into non-overlapping 10-second windows, and each window is processed with three digital filters. The first filter removes baseline wander and high-frequency noise in preparation for spectral HR estimation. This HR estimate is then used to determine the upper cutoff frequency of the second and third filters. These filters help to detect peaks in the PPG signal and its first derivative (PPG') respectively, and ultimately determine the systolic peaks. The peak detection threshold is set at the 90$^{th}$ percentile of the amplitude of the PPG', signal and the 60$^{th}$ percentile of the PPG signal. The systolic peaks are identified as peaks in a weakly filtered PPG signal immediately following each peak identified in the differentiated signal \cite{charlton_2022detecting}.

%%%%%%%%%%

\subsection{\textit{Aboy+} peak detector}

The \textit{Aboy+} algorithm is an accelerated version of the original \textit{Aboy} algorithm \cite{Aboy2005}. Charlton's implementation of the \textit{Aboy} algorithm in Matlab \cite{charlton_2022detecting} uses high-order finite impulse response (FIR) filters for the three bandpass filters with passbands of 0.45–10 Hz, 0.45–2.5*HR/60 Hz, and 0.45–10HR/60 Hz. Although this provides accurate cutoff frequencies, the filters are computationally expensive, particularly in the case of long-term measurements (see Table \ref{tab_02}). Therefore, in the Aboy+ algorithm, zero-phase 5$^{th}$ order Chebyshev Type II, infinite impulse response (IIR) filters with the same cutoff frequencies, were used to reduce the computational complexity.

%%%%%%%%%%

\subsection{\textit{Aboy++} peak detector}

\textit{Aboy++} peak detection algorithm includes adaptive HR estimation, to improve \textit{Aboy+}. In each 10-second window, an HR estimate is made for the subsequent window using constraints to reduce the likelihood of incorrect HR values. Based on the estimated HR, the upper cutoff frequency is adjusted to improve peak detection. In summary the \textit{Aboy++} algorithm involves the following steps: 

\begin{enumerate}
     \item \textbf{Windowing}: 10 s, non-overlapped windows are used. 
     \item \textbf{HR estimation}: The HR estimation process utilizes the \textit{DetectMaxima} function, which is enhanced through the implementation of adaptive peak-to-peak time and a modified percentile of peak amplitude. % (see Algo. \ref{algo_DetectMaxima}).
     \item \textbf{Define HR index}: The list of the peak-to-peak times above 30$^{th}$ percentile is defined as $P_d$. If $median(P_d)*0.5<mean(P_d)<mean(P_d)*1.5$, then $HR_i=std(P_d)/mean(P_d)*10$. Otherwise, the $HR_i$ is the same as in the previous window.
     \item \textbf{Define HR window}: $HR_{win}=F_s/((1+HR_i)*3)$,\\ where $F_s$ is the sampling frequency.
     \item \textbf{Final peaks}: Using \textit{DetectMaxima}, the peak-to-peak time must be at least $2*HR_{win}$ and prominence must be at least 25\% of the average systolic peak amplitude.
     \item \textbf{Update HR window}: for the subsequent 10 s window. 
 \end{enumerate}
 
\vspace{.1cm}
\noindent The open-source algorithm is available on the physiozoo.com website.

\begin{figure*}[!hbt]
  \centering
  \includegraphics[width=.97\textwidth]{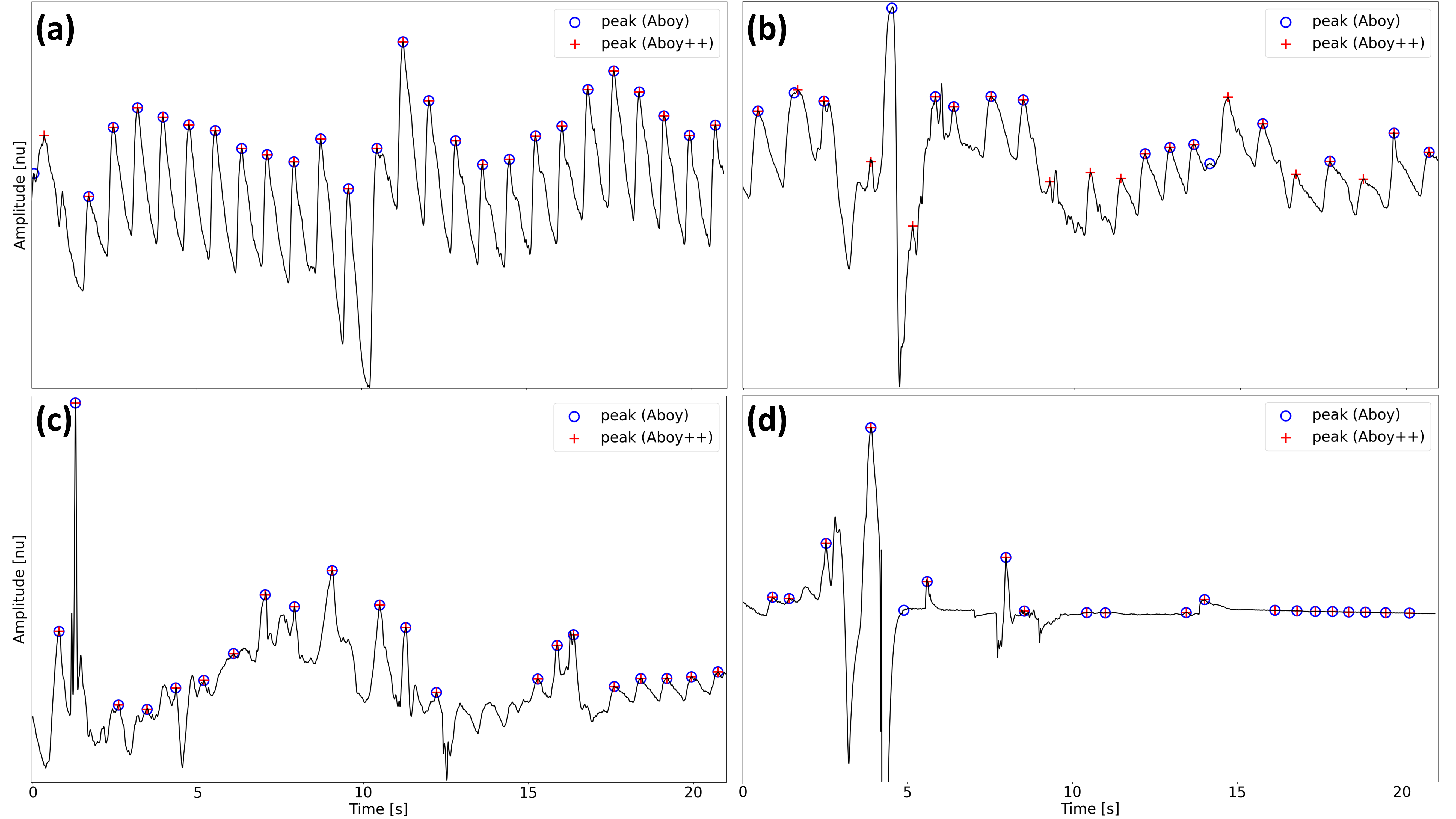}
  \caption{This figure illustrates the performance of the \textit{Aboy} and \textit{Aboy++} peak detectors on PPG signals of varying quality. Panel (a) represents a good-quality PPG signal with slight baseline wander, while Panel (b) represents a signal with rapid amplitude fluctuations. Panel (c) displays a low-quality PPG signal, while Panel (d) demonstrates very low-quality PPG signals of signals in the long term.}
  \label{fig:comp_peak_detection}
\end{figure*}

\section{Materials and Methods}

%%%%%%%%%%
\subsection{MESA database}

The MESA database is a well-known database widely used cardiovascular disease-related research \cite{Dean_2016}. It consists of data from 2,054 adult individuals with sub-clinical cardiovascular disease and contains a total of 19,998 hours of PPG records. The male and female ratio is 1:1.2, and the patients age range is 54-95 years. The database was downloaded from the National Sleep Research Resource (NSRR). The polysomnography (PSG) records were acquired at home from the fingertip using Nonin ® 8000 series pulse oximeters (Nonin Medical Inc.). The PPG sampling rate was 256 Hz. 

%\vspace{-.2cm}
\subsection{Evaluation of peaks}
%\vspace{-.2cm}

To forecast the PPG systolic peaks, the electrocardiogram (ECG) peaks were extracted from the PSG records and used as a reference \cite{Kotzen_2021}. The Kotzen \textit{et al.} \cite{Kotzen_2021} Peak Matching method was used to benchmark the \textit{Aboy}, \textit{Aboy+}, and \textit{Aboy++} algorithms on the MESA database. However, Kotzen \cite{Kotzen_2022_MSc} excluded noisy signals during evaluation, which resulted in higher F1-scores for the previously benchmarked peak detectors than in the present study.

The original and improved algorithms were evaluated on a standalone server using 100 records and 1025 hours of PPG data, from the MESA database. The median length of the records was 10 hours, with an interquartile range (IQR) of 2 hours. The records were divided into 1-hour-long segments, and the F1-score was calculated as the mean value of these segments.

To present the F1-score results, the median (MED) and the quartiles (Q1, Q3) were applied. MED: Value in the middle of the ordered set of data. Q1, Q3: The data value at which percent of the value in the data set is less than or equal to this value, calculated with 25$^{th}$ or 75$^{th}$.

\section{Results}
%\vspace{-.2cm}
%\subsection{PPG peak detection}\label{results_peak}
%\vspace{-.2cm}
Table \ref{tab_01} shows that \textit{Aboy++} outperformed the original \textit{Aboy} by almost  5\% in terms of F1-score. Our initial expectation was that \textit{Aboy} and \textit{Aboy+} would achieve similar F1 scores. However, we found that \textit{Aboy+} significantly outperformed \textit{Aboy}. Although both \textit{Aboy} and \textit{Aboy+} used the same cutoff frequency, the 4$^{th}$ order IIR filter used in \textit{Aboy+} was more effective in processing low-quality signals compared to the FIR filter in the original \textit{Aboy} implemented by Charlton \cite{charlton_2022detecting}. The median length of each patient's recording was 10 hours, with an interquartile range of 2 hours. When evaluated on recordings of 100 adults included the MESA database containing more than 4.25 million reference beats, \textit{Aboy++} achieved an F1-score of 85.5\%, compared to 80.99\% for the \textit{Aboy+} peak detector.

\begin{table}[h]
\fontsize{9pt}{12pt}
\selectfont
\begin{center}
\caption{F1-score of peak detectors}\label{tab_01}%
\begin{tabular}{l|l|l|l|}

\cline{1-4}
\multicolumn{1}{|c|}{\multirow{2}{*}{\textit{\textbf{\begin{tabular}[c]{@{}c@{}}100\\ records\end{tabular}}}}} & \multicolumn{3}{c|}{\textbf{F1-score}}              
\\ \cline{2-4} \multicolumn{1}{|c|}{} & \multicolumn{1}{l|}{\textbf{Aboy}} & \textbf{Aboy+} & \multicolumn{1}{l|}{\textbf{Aboy++}} \\ \hline
\multicolumn{1}{|l|}{\textit{MED}} & 80.99\% & 84.62\% & 85.50\%\\ \hline
\multicolumn{1}{|l|}{\textit{Q1}} & 73.91\% & 78.63\% & 79.78\%\\ \hline
\multicolumn{1}{|l|}{\textit{Q3}} & 85.52\% & 90.21\% & 92.57\%\\ \hline
\end{tabular}
\end{center}
\end{table}

The open-source implementation of the original \textit{Aboy} peak detector \cite{charlton_2022detecting}, was not efficient for long-term measurement analysis due to its high computational cost. The enhanced performance of the \textit{Aboy++} peak detector was not limited to its superior F1-score. \textit{Aboy++} also provided a remarkable reduction in computational time compared to the original \textit{Aboy} algorithm, as summarized in Table \ref{tab_02}. The processing time for detecting peaks using \textit{Aboy++} was over 57 times faster than the original algorithm without multiprocessing. This significant improvement in speed can be particularly advantageous for long-term PPG recordings, where the original \textit{Aboy} algorithm was computationally expensive.

\begin{table}[h]
\fontsize{9pt}{12pt}
\selectfont
\begin{center}
\caption{Computational time of peak detectors}\label{tab_02}%
\begin{tabular}{|l|ll|ll|}
\hline
\multicolumn{1}{|c|}{\multirow{2}{*}{\textit{\textbf{\begin{tabular}[c]{@{}c@{}}100\\ records\end{tabular}}}}} & \multicolumn{2}{c|}{\textbf{1 hour}}                 & \multicolumn{2}{c|}{\textbf{10   hour}}              \\ \cline{2-5} 
\multicolumn{1}{|c|}{} & \multicolumn{1}{l|}{\textbf{Aboy}} & \textbf{Aboy++} & \multicolumn{1}{l|}{\textbf{Aboy}} & \textbf{Aboy++} \\ \hline
\textit{MED} & \multicolumn{1}{l|}{114.24 sec} & 1.98 sec & \multicolumn{1}{l|}{1116.72 sec}   & 19.67 sec       \\ \hline
\textit{IQR} & \multicolumn{1}{l|}{8.28 sec} & 0.14 sec & \multicolumn{1}{l|}{242.10 sec}    & 4.21 sec        \\ \hline
\end{tabular}
\end{center}
\end{table}

\vspace{-.8cm}
\section{Conclusion and discussion}
\vspace{-.3cm}
This research contributes a new PPG peak detector denoted \textit{Aboy++} which was based on the original \textit{Aboy} algorithm developed by Aboy \textit{et al.} The adaptive feature of the \textit{Aboy++} peak detector provides higher accuracy compared to the original \textit{Aboy} peak detector in cases where there is strong baseline wander (see Fig. \ref{fig:comp_peak_detection}.b), rapid amplitude fluctuations, and high heart rate variability. In the presence of a noisy signal (see Fig. \ref{fig:comp_peak_detection}.b) peak detection can also be challenging for the \textit{Aboy++} peak detector. In the short term, the estimated heart rate of the previous windows can provide a reasonable output (see Fig. \ref{fig:comp_peak_detection}.c); however, peak detection accuracy can be diminished in the presence of long-term noisy signals (see Fig. \ref{fig:comp_peak_detection}.d). Although excluding low-quality signals can have advantages, inaccurate assessment of signal quality can lead to significant changes in the overall signal evaluation.

The performance of \textit{Aboy++} has shown superior performance compared to other open-source algorithms benchmarked on the MESA database, making it a strong candidate for a standardized toolbox for comprehensive PPG analysis. These results demonstrate that \textit{Aboy++} is a highly accurate and efficient peak detection algorithm that can be reliably used for PPG signal analysis. Future work will involve evaluating the performance of Aboy++ on the complete MESA database, as well as on other databases of PPG signals from different patient groups. Additionally, there will be an implementation of algorithms for detecting other PPG fiducial points, such as the pulse onset, dicrotic notch, diastolic peak, and fiducial points of higher-order derivatives, which rely heavily on the accurate identification of systolic peaks. By using \textit{Aboy++} as a foundation for PPG analysis, researchers can improve the accuracy and reliability of their findings and make significant progress in understanding the cardiovascular system's underlying mechanisms.

\vspace{-.3cm}
\section{Acknowledgements}
\vspace{-.3cm}
MG and JB acknowledge the Estate of Zofia (Sophie) Fridman and funding from the Israel Innovation Authority. PHC acknowledges funding from the British Heart Foundation [FS/20/20/34626].

The Multi-Ethnic Study of Atherosclerosis (MESA) Sleep Ancillary study was funded by NIH-NHLBI Association of Sleep Disorders with Cardiovascular Health Across Ethnic Groups (RO1 HL098433). MESA is supported by NHLBI funded contracts HHSN268201500003I, N01-HC-95159, N01-HC-95160, N01-HC-95161, N01-HC-95162, N01-HC-95163, N01-HC-95164, N01-HC-95165, N01-HC-95166, N01-HC-95167, N01-HC-95168 and N01-HC-95169 from the National Heart, Lung, and Blood Institute, and by cooperative agreements UL1-TR-000040, UL1-TR-001079, and UL1-TR-001420 funded by NCATS. The National Sleep Research Resource was supported by the National Heart, Lung, and Blood Institute (R24 HL114473, 75N92019R002).

% \section{Supplementary Material}
% \vspace{-.2cm}
% \begin{algorithm}[!hbt]
% \fontsize{9}{9}
% \caption{Modified DetectMaxima}\label{algo_DetectMaxima}
% \begin{program}
% \text{Inputs}
% sig := \text{Input signal.}
% hr_{win} := \text{HR. window.}
% prctile_x := \text{percentile at x\%.}

% \text{Output}
% max_{pks} := \text{Detected peaks}.

%     \BEGIN \\
%         s1 = sig(3:end), 
%         s2,s3 = sig(2:end-1), 
%         sig(1:end-2)
                
%         \IF hr_{win} == 0,
%             m = 1 + find((s1 < s2) and (s3 < s2))
%             max_{pks} = m(sig(m) > tr)
%         else
%                 distance=hr_{win}
%             max_{loc} = findpeaks(sig, distance)
%             min_{loc} = findpeaks(-sig, distance)
        
%             \FOR i:= 0 \TO length(max_{loc}),
%                 values = abs(max_{loc}(i) - min_{loc})
%                 min_v, min_i = minimum(values)
%                 intensity.append(sig(max_{loc}(i)) - sig(min_{loc}(min_i)))
%             \END
        
%             prominence = mean(intensity)*0.25
%                 distance=hr_{win}
%             min_v, min_i= minimum(sig)
%                 sig_{new}=sig+min_v
%             max_{pks} = findpeaks(sig_{new},prominence,distance)
%         \END
        
%         \text{return } max_{pks}
%     \END
% \end{program}
% \end{algorithm}

\bibliographystyle{cinc}
\vspace{-.2cm}
\bibliography{new_refs}
\vspace{-.7cm}

\begin{correspondence}
\vspace{-.2cm}
Márton Á. Goda, PhD\\
marton.goda@campus.technion.ac.il\\
AIMLab, Technion-IIT, Haifa, 32000, Israel\\
\end{correspondence}
\vspace{-3cm}
\end{document}